\documentclass[sigconf]{acmart}
\AtBeginDocument{%
  }
\copyrightyear{2024}
\acmYear{2024}
\setcopyright{rightsretained}
\acmConference[CIKM '24]{Proceedings of the 33rd ACM International Conference on Information and Knowledge Management}{October 21--25, 2024}{Boise, ID, USA}
\acmBooktitle{Proceedings of the 33rd ACM International Conference on Information and Knowledge Management (CIKM '24), October 21--25, 2024, Boise, ID, USA}
\acmDOI{10.1145/3627673.3679987}
\acmISBN{979-8-4007-0436-9/24/10}

\usepackage{bm}
\usepackage{caption}
\usepackage{subcaption}
\usepackage{algorithm}
\usepackage{algorithmic}
\usepackage[flushleft]{threeparttable}
\usepackage{amsmath}
\usepackage{multirow}
\usepackage{tabularx}
\usepackage{array}
\usepackage{xcolor}
\usepackage{balance}
\begin{document}

\title{RecPrompt: A Self-tuning Prompting Framework for News Recommendation Using Large Language Models}
% \titlerunning{RecPrompt: A Prompt Engineering Framework}

\author{Dairui Liu}
% \authornote{First author}
\orcid{0000-0002-8573-3857}
\affiliation{%
  \institution{University College Dublin}
  \city{Dublin}
  \country{Ireland}
}
\email{dairui.liu@ucdconnect.ie}

\author{Boming Yang}
\orcid{0009-0004-6298-5711}
\affiliation{
  \institution{The University of Tokyo}
  \city{Tokyo}
  \country{Japan}
}
\email{boming.yang@weblab.t.u-tokyo.ac.jp}

\author{Honghui Du}
\orcid{0000-0001-8758-0092}
\affiliation{%
  \institution{University College Dublin}
  \city{Dublin}
  \country{Ireland}
}
\email{honghui.du@insight-centre.org}

\author{Derek Greene}
\author{Neil Hurley}
% \orcid{0000-0001-8065-5418}
\affiliation{%
  \institution{University College Dublin}
  \city{Dublin}
  \country{Ireland}
}
% \email{derek.greene@ucd.ie}
% \email{neil.hurley@ucd.ie}

\author{Aonghus Lawlor}
\author{Ruihai Dong}
% \orcid{0000-0001-8428-2866}
\affiliation{%
  \institution{University College Dublin}
  \city{Dublin}
  \country{Ireland}
}
% \email{aonghus.lawlor@ucd.ie}
% \email{ruihai.dong@ucd.ie}

\author{Irene Li}
\orcid{0000-0002-1851-5390}
\affiliation{
  \institution{University of Tokyo}
  \city{Tokyo}
  \country{Japan}
}
\email{ireneli@ds.itc.u-tokyo.ac.jp}

% \renewcommand{\shortauthors}{Liu et al.}
% Requested by the conference.
\renewcommand{\shortauthors}{Dairui Liu et al.}
\begin{abstract}
News recommendations heavily rely on Natural Language Processing (NLP) methods to analyze, understand, and categorize content, enabling personalized suggestions based on user interests and reading behaviors. Large Language Models (LLMs) like GPT-4 have shown promising performance in understanding natural language. However, the extent of their applicability to news recommendation systems remains to be validated. This paper introduces RecPrompt\footnote{Source code:\url{https://github.com/Ruixinhua/rec-prompt}. This work was supported and funded by the Science Foundation Ireland through the Insight Centre for Data Analytics (Grant no. SFI/12/RC/2289\_P2), EU Horizon Europe SEDIMARK (SEcure Decentralised Intelligent Data MARKetplace) project (Grant no. 101070074), and the Japan Society for the Promotion of Science (JSPS) KAKENHI (Grant no. 24K20832).}, the first self-tuning prompting framework for news recommendation, leveraging the capabilities of LLMs to perform complex news recommendation tasks. This framework incorporates a news recommender and a prompt optimizer that applies an iterative bootstrapping process to enhance recommendations through automatic prompt engineering. Extensive experimental results with 400 users show that RecPrompt can achieve an improvement of 3.36\% in AUC, 10.49\% in MRR, 9.64\% in nDCG@5, and 6.20\% in nDCG@10 compared to deep neural models. Additionally, we introduce TopicScore, a novel metric to assess explainability by evaluating LLM's ability to summarize topics of interest for users. The results show LLM's effectiveness in accurately identifying topics of interest and delivering comprehensive topic-based explanations.
\end{abstract}





  
\maketitle
% \vspace{-3pt}
\section{Introduction}
\label{sec:introduction}
The rise of personalized news recommendation systems, like Google News and Microsoft News, has revolutionized how people access and consume news by making it more tailored to individual preferences \cite{wu2023personalized}. Achieving high-quality news recommendations necessitates precisely understanding the semantics within news articles \cite{xiao2022training}. Recent advancements in natural language processing, particularly through Large Language Models (LLMs) such as GPT-4~\cite{openai2023gpt4}, have shown considerable potential in this area. These models leverage extensive linguistic and world knowledge from large-scale corpora to understand and generate human-like language, making them valuable for aligning news content with user preferences and enhancing explainability in recommendation systems \cite{Fan2023RecommenderSI}.

Despite the promise of LLMs, current implementations in news recommendation systems face challenges. Initial methods by integrating LLMs into conventional deep neural models~\cite{Liu2023ONCEBC,runfeng2023lkpnr} have shown good recommendation performance but often lose the advantages of LLMs. 
Fine-tuning LLMs~\cite{Li2023ExploringFC,AliHusseinAlNaffakh2024ExploringCP} is a way to retain these advantages, but it requires large sets of high-quality rationales, which are complex and resource-intensive to produce~\cite{wei2022chain}. 
An alternative approach is to leverage the inherent capabilities of LLMs using prompts—text templates to guide model responses~\cite{Xu2024PromptingLL,Yang2024KGRankEL} with different prompting strategies. For instance, Dai et al.~\cite{Dai2023UncoveringCC} explore using LLMs with a simple input-output (IO) prompting strategy for direct textual responses in recommendations. RecRanker~\cite{Luo2023RecRankerIT} enhances the prompt through instruction tuning with auxiliary information from recommendation models. LLM-Rec~\cite{Lyu2023LLMRecPR} employs diverse prompting strategies with LLM-augmented text to enhance recommendations. These prompting strategies have demonstrated effectiveness in complex tasks but do not surpass well-trained deep neural methods~\cite{LSTUR2019MingxiaoAn,NAML2019ChuhanWu, NRMS2019ChuhanWu,NPA2019ChuhanWu,DKN2018Wang,glory23,liu2023topiccentric} in performance. Additionally, they require significant human effort for tuning and evaluation~\cite{liu2023pre,Fan2023RecommenderSI,Gao2023LargeLM,Song2023NLPBenchEL}.
The nature of LLMs makes them well-suited for generating explanations for recommendations. For example, ChatRec~\cite{gao2023chat} adopts pre-trained LLMs to provide more interactive and explainable recommendations. However, evaluating recommender systems has primarily focused on ranking-based metrics~\cite{wu-2020-mind}. Few works focus on studying the generation of explanations for recommendations and their evaluation due to the lack of ground truth. 

To overcome these limitations, we introduce \textbf{RecPrompt}, a self-tuning prompting framework specifically designed for news recommendations. RecPrompt operates through an iterative bootstrapping process involving two LLMs: a news recommender and a prompt optimizer. The recommender generates recommendations with text prompts, and these are then fed into the optimizer, which refines the prompt based on the instructions, enhancing the prompt's alignment with user preferences and topics of interest. Additionally, we propose \textbf{TopicScore}, a new metric addressing the shortcomings of existing metrics by focusing on the semantic relevance of topics. Extensive experiments demonstrate significant improvements in recommendation performance with an increment of 3.36\% in AUC, 10.49\% in MRR, 9.64\% in nDCG@5, and 6.20\% in nDCG@10 over traditional deep neural methods and validate RecPrompt’s superiority in generating relevant topics.

\begin{figure}[!t]
    \centering
    \includegraphics[width=\linewidth,height=3.6cm]{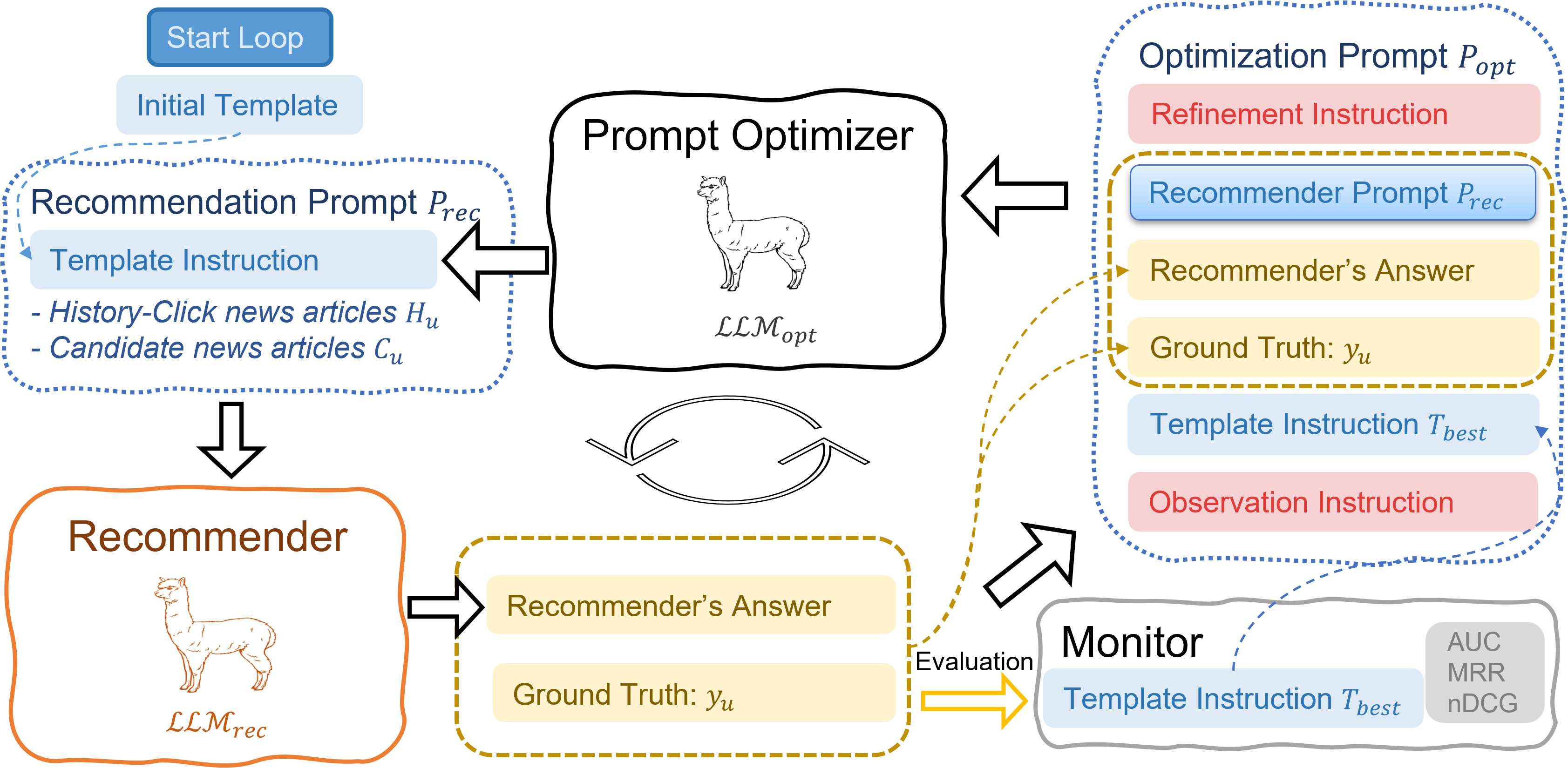}
    \caption{\textbf{RecPrompt: a self-tuning prompting Framework.}}
    \label{fig:RecPrompt_pipeline}
\end{figure}
\section{Proposed Method: RecPrompt}
The goal of news recommendation is to predict a list of candidate news articles for users based on their reading history. We define the history for a user \( u \) as \( \bm{H}_u = \{nw_i\} \), where each news article \( nw \) includes a title and a category. The candidate news is defined as \( \bm{D}_u = \{(nw_j, y_j)\} \), with each news article accompanied by a label \( y_j \in \{0,1\} \). A value of 1 indicates that the user is interested in the candidate news, while 0 indicates no interest. The objective is to learn a news recommender, and for any unseen user $u'$, the output is a ranked list of the candidate news $\bm{R}_{u'}$. 

We propose the RecPrompt framework to achieve this recommendation objective by improving the prompt to enhance the LLM's ability to summarize topics of the user's interests, thus improving recommendations performance. As shown in Figure~\ref{fig:RecPrompt_pipeline}, RecPrompt consists of three main components: a \textbf{News Recommender} ($\bm{LLM}_{rec}$), a \textbf{Prompt Optimizer} ($\bm{LLM}_{opt}$), and a \textbf{Monitor}. 
The LLM-based recommender takes recommendation prompts $P_{rec}$ as inputs and outputs not only a ranked list $\bm{R}_u$ of candidate news articles but also explanations summarizing the topics $\bm{TP}_u$ of interest based on user reading history.
The monitor evaluates recommendations and records the best template instruction for the prompt optimizer. Also, the prediction of the news recommender and the recommendation prompt, $P_{rec}$, are then fed into the LLM-based Prompt Optimizer with an optimization prompt to generate a refined template instruction used to form next-round recommendation prompts.

\vspace{-2pt}
\subsection{RecPrompt Components}

\subsubsection{News Recommender}
\label{sec: Initializing Recommender Prompt}
We utilize an LLM to make recommendations, so the recommendation prompt, $P_{rec}$  is required. This prompt contains a template instruction $T$, as well as two input placeholders,``\$\{history\}'' and ``\$\{candidate\}'', for $\bm{H}_u$ and $\bm{C}_u$ for each user. This template can be initialized by any prompting strategy, such as Input-Output (IO) prompting or Chain of Thoughts~\cite{kojima2022large} prompting. An example of using the IO-prompting template is:
\vspace{-2pt}
\begin{verbatim}
You serve as a personalized news recommendation system.
# Input Format
## User's History News
${history}
## Candidate News
${candidate}
# Output Format
Rank candidate news based on the user’s history news in 
the format: "Ranked news: <START>C#, C#,..., C#<END>".
\end{verbatim}
\vspace{-2pt}
\noindent The output of the news recommender contains a ranked news sequence ($\bm{R}_u$) and explanation ($TP_u$). The explanation is a sequence of preference topics summarized from the user's reading history and the corresponding news list for each topic. 

\subsubsection{Prompt Optimizer}The task of Prompt Optimizer is to generate a better recommendation prompt. We utilize another LLM to achieve this by taking an optimization prompt as input. This prompt contains a refinement instruction, the recommendation prompt ($P_{rec}$) for a user $u$, the ranked news sequence ($\bm{R}_u$) and the explanation ($TP_u$), as well as the ground truth ($y_u$), best template recorded by the Monitor, and an observation instruction. The refinement and observation instructions are crucial in guiding the optimizer to a specific direction. 
The refinement instruction typically starts with a sentence like "\textit{\textbf{You should generate an improved template instruction based on the provided information.}}" at the beginning.
The observation instruction is designed to focus on topics for the news recommendation task. An example observation instruction could be:

\noindent\texttt{You should focus on how well the recommender's response aligns with the user's click behavior by examining if the \underline{topics} from the user's news history and candidate news are accurately summarized and matched to the user's interests. Specifically, evaluate the clarity of \underline{topics} in the recommender's answer, review the task description for adequacy, and check the detail in the recommendation process to ensure it reflects an analysis and summary of user-interest \underline{topics}.}

\noindent These instructions direct the optimizer to effectively summarize the topics of news articles to aid in making accurate recommendations and explanations with topics. The optimizer uses the optimization prompt to update the current recommendation prompt, which can be used for next-round recommendations.

\subsubsection{Monitor}

It evaluates the recommender's performance using the templates generated by the optimizer. It observes recommendation performance by computing metrics such as MRR (Mean Reciprocal Rank) and nDCG (normalized Discounted Cumulative Gain)~\cite{wu-2020-mind} to determine whether the current template represents an improvement over previous ones. These metrics are calculated based on the ranked list $\bm{R}_u$ and clicked ground truth $y_u$ for all users. By continuously assessing these metrics, the Monitor ensures that the most effective template is identified and retained for further use, thereby enhancing the overall performance of the news recommendation system.

\subsection{Tuning Procedure}
As shown in Figure~\ref{fig:RecPrompt_pipeline}, the recommender and optimizer operate in a loop. The process starts with an initial template filled with specific data to form the complete recommendation prompt to start the iteration loop. The recommender LLM generates recommendations based on this initial template, using the user's history of clicked news articles and a set of candidate news articles.
After the recommender generates its recommendations, the optimizer updates the current template. This refined template is then provided to the recommender in the next iteration to make a new prediction. The template will only be updated if the Monitor evaluates and confirms a performance improvement. The iteration loop continues until a specified number of iterations ($l$) is reached, ensuring continuous enhancement of the recommendation system.

\subsection{Evaluating Explanation: TopicScore}
\label{sec:topic_score}
RecPrompt generates enhanced prompts that enable $\bm{LLM}_{rec}$ to explain its recommendations by summarizing topics of interest~\cite{liu2023topiccentric,BATM2022Liu}. 
The explanation is a list of topics summarizing the user's reading history and a sequence of news articles that are classified to the specific topic. 
For example, a topic of interest for the user is \textit{sports}, and the related history news articles are \textit{H1, H3}. A good explanation should correctly recognize user interest on the topics, and accurately classify the history news articles. 
The existing methods like BERTScore~\cite{zhang2019bertscore} are inadequate to measure how an explanation can correctly recognize topics of user's interest, so we propose TopicScore to assess these topics, which evaluates the correctness and completeness of topics within explanations. To evaluate TopicScore, we use LLM evaluator $\bm{LLM}_{eval}$ and human annotations to provide ground truth for scoring. 

\noindent{\textbf{Correctness}} measures how accurately the generated topics of $\bm{LLM}{rec}$ reflect the news articles: 
\begin{equation} 
TS_{correctness} = \frac{\sum_{u \in U} \sum_{tp \in TP_u} \mathbb{I}(tp)}{\sum_{u \in U} |TP_u|} 
\end{equation} 
where $\mathbb{I}(\cdot)$ is an indicator function that equals 1 if the topic $tp$ matches the corresponding news article and 0 otherwise, and $|TP_u|$ is the number of summarized topics for the user.

\noindent{\textbf{Completeness}} evaluates how well the summarized topics cover the user's interests according to historical records: 
\begin{equation} 
TS_{completeness} = \frac{\sum_{u \in U} |H_{m}^{u}|}{\sum_{u \in U} |H^{u}|} 
\end{equation} 
where $|H_{m}^{u}|$ is the number of history news articles covered by the topics for user $u$, $|H^{u}|$ is the number of historical clicks for user $u$.

\noindent{\textbf{LLM Evaluator.}} We collect a set of topic summaries generated by $\bm{LLM}_{rec}$ for various news articles. The LLM evaluator, $\bm{LLM}_{eval}$, uses an evaluation prompt to check if each summarized topic accurately reflects the content of the corresponding news article. 

\noindent{\textbf{Human annotation.}} Mirroring the evaluation by $\bm{LLM}_{eval}$, we establish a workflow for human annotators to select all relevant topics that correctly match the given news content.

\section{Experiments}
\subsection{Experimental Settings}
We evaluate our proposed model on a benchmark news recommendation dataset, the MIcrosoft News Dataset (MIND) collection~\cite{wu-2020-mind}. From MIND, we randomly select 100 users to form the validation set for optimizing RecPrompt and 400 users to constitute the test set. 
%Each record includes one positively clicked news item by the user and nine negative samples (i.e., news items not clicked) to simulate the candidate news set. 
Each user is given 10 candidate news articles that are shown in an impression and only one of them is clicked by the user, where each news item includes headlines and its category tag. All experiments are conducted 3 times on the test set, with the average performance metrics reported to ensure consistency. 
The recommender component is represented as $\bm{LLM}_{rec}$, the optimizer component as $\bm{LLM}_{opt}$, and the evaluator component as $\bm{LLM}_{eval}$. Each component can operate with either GPT-3.5 or GPT-4, indicated by subscripts \text{\textbf{-3.5}} and \text{\textbf{-4}} respectively. We limit the number of iterations $l$ for RecPrompt to 10 and apply a zero-shot prompt \cite{Sileo2021ZeroShotRA} for $\bm{LLM}_{rec}$. We utilize OpenAI's API versions gpt-3.5-turbo-1106 for GPT-3.5 and gpt-4-1106-preview for GPT-4, respectively, applying for $\bm{LLM}_{rec}$, $\bm{LLM}_{opt}$, and $\bm{LLM}_{eval}$. Following previous work~\cite{wu-2020-mind}, we use the same pipeline to train deep neural models and consider four metrics to evaluate the performance of news recommendations: AUC, MRR, nDCG@5 (N@5), and N@10. We apply two prompting strategies, Input-Output (IO) \cite{Dai2023UncoveringCC} and Chain of Thoughts (CoT) \cite{kojima2022large} for the template. 
\newcolumntype{C}[1]{>{\centering\arraybackslash}m{#1}}

\begin{table}[!b]
    \centering
    \caption{Recommendation performance of models regarding AUC, MRR, nDCG@5, and nDCG@10. 
    % The best results are highlighted and the underscore marks the best result in the deep neural news recommendation model group. 
    }
    \begin{tabular}{@{} p{2.15cm}@{}C{1.25cm}C{1.25cm}C{1.25cm}C{1.25cm}}
    \hline
    \textbf{Model} &  AUC & MRR & nDCG@5 & nDCG@10 \\
    \hline
    Random & 50.89±2.23 & 30.30±1.17 & 30.47±1.99 & 46.22±0.99 \\
    MostPop & 52.47±0.04 & 34.99±0.01 & 34.68±0.05 & 49.77±0.01  \\
    TopicPop & 64.64±0.04 & 39.35±0.07 & 44.39±0.08 & 53.59±0.05 \\
    \hline
    Deep Model & & & & \\
    LSTUR~\cite{LSTUR2019MingxiaoAn} & \underline{67.17±0.22} & 43.76±0.30 & 47.84±0.23 & 57.00±0.23 \\
    DKN~\cite{DKN2018Wang}  & 66.28±0.43 & 42.31±0.50 & 46.43±0.37 & 55.88±0.38 \\
    NAML~\cite{NAML2019ChuhanWu} & 66.79±0.10 & \underline{44.03±0.46} & \underline{48.03±0.35} & \underline{57.18±0.34} \\
    NPA~\cite{NPA2019ChuhanWu} & 65.52±0.78 & 43.16±0.40 & 46.53±0.47 & 56.47±0.32 \\
    NRMS~\cite{NRMS2019ChuhanWu} & 66.25±0.12 & 43.60±0.30 & 46.86±0.25 & 56.82±0.21 \\
    \hline
    Prompt LLM & & & & \\
    IO-$\bm{LLM}_{rec\text{-}3.5}$ & 59.08±0.23 & 38.55±0.15 & 40.35±0.38 & 52.74±0.21 \\
    CoT-$\bm{LLM}_{rec\text{-}3.5}$ & 58.67±0.14 & 37.66±0.35 & 39.12±0.26 & 52.06±0.41 \\
    IO-$\bm{LLM}_{rec\text{-}4}$ & 65.06±0.14 & 43.62±0.13 & 46.66±0.24 & 56.78±0.06 \\
    CoT-$\bm{LLM}_{rec\text{-}4}$ & 66.06±0.18 & 44.01±0.29 & 48.02±0.34 & 57.12±0.17  \\
    \hline
    RecPrompt & & & & \\
    IO-$\bm{LLM}_{rec\text{-}3.5}$ & 62.28±0.14 & 39.53±0.12 & 43.65±0.32 & 53.65±0.25 \\
    CoT-$\bm{LLM}_{rec\text{-}3.5}$ & 64.73±0.32 & 42.22±0.23 & 46.33±0.11 & 55.75±0.19 \\
    IO-$\bm{LLM}_{rec\text{-}4}$ & 69.39±0.21 & 48.2±0.34 & 52.01±0.37 & 60.39±0.29 \\
    CoT-$\bm{LLM}_{rec\text{-}4}$ & \textbf{69.43±0.04} & \textbf{48.65±0.32} & \textbf{52.66±0.43} & \textbf{60.73±0.24}  \\
    \hline
    Increment->DM & (+3.36\%) & (+10.49\%) & (+9.64\%) & (+6.20\%) \\
    \hline
    \end{tabular}
    \label{tab:over_results}
   % \vspace{-15pt}
\end{table}
\subsection{Results and Discussion}
Table~\ref{tab:over_results} summarizes the results of all methods on the test dataset, leading to several key observations. Firstly, among traditional baselines, TopicPop achieves the best performance. This success can be attributed to its use of category tags from the news data, which provide more tailored recommendations. This demonstrates the effectiveness of leveraging topical information to match user interests. Secondly, all deep neural models outperform TopicPop. These models benefit from their advanced capabilities in processing and representing the textual content of news articles. In particular, NAML outperforms other baselines because it extracts comprehensive information from news text, categories, and subcategories, resulting in more informative representations. In terms of the impact of the prompting strategy, it shows a clear advantage of the CoT strategy over IO in improving recommendation performance across most configurations. Regarding our proposed RecPrompt methods, the results indicate that CoT-$\bm{LLM}_{rec\text{-}3.5}$ outperforms all traditional methods. Notably, it slightly surpasses TopicPop, illustrating that $\bm{LLM}_{rec\text{-}3.5}$ is more effective in summarizing user topics from headlines and category tags. Lastly, CoT-$\bm{LLM}_{rec\text{-}4}$ achieves superior results compared to all deep neural models using zero-shot prompting, without any training on news recommendation data. This highlights RecPrompt's ability to effectively understand news content and provide accurate recommendations.

\subsection{Ablation Study}
We systematically examine the impact of key components within the RecPrompt framework to understand their contributions to enhancing the performance of news recommendation systems. The study focuses on the use of different LLMs as the recommender and the optimizer, and the inclusion of category information in the model inputs. The performance outcomes are detailed in Table \ref{tab:ablation_study}.
\begin{table}[!b]
\centering
\caption{Performance comparison of $\bm{LLM}_{rec}$ and $\bm{LLM}_{opt}$ on RecPrompt, considering various settings.
}
\begin{tabular}{@{}p{2.5cm}@{}C{0.6cm}C{0.6cm}p{0.6cm}C{0.6cm}C{0.6cm}p{0.6cm}}
\hline
\multirow{2}{*}{Strategy} & \multicolumn{3}{c}{w/o Category} & \multicolumn{3}{c}{with Category} \\ \cline{2-7}
             & AUC    & MRR    & N@5 & AUC   & MRR   & N@5 \\ \hline
$\bm{LLM}_{rec\text{-}3.5}$      &   &  &   &   &   &       \\
IO-$\bm{LLM}_{opt\text{-}3.5}$  & 58.97 & 35.85 & 39.17 & 61.24 & 39.23 & 41.97 \\
CoT-$\bm{LLM}_{opt\text{-}3.5}$ & 58.96 & 37.46 & 39.62 & 64.06 & 41.26 & 44.55  \\ 
IO-$\bm{LLM}_{opt\text{-}4}$    & 59.32 & 36.15 & 39.42 & 62.28 & 39.53 & 43.65 \\ 
CoT-$\bm{LLM}_{opt\text{-}4}$   & \underline{60.03} & \underline{38.18} & \underline{40.89} & \underline{64.73} & \underline{42.22} & \underline{46.33} \\
\hline
$\bm{LLM}_{rec\text{-}4}$         &   &  &   &   &   &  \\
IO-$\bm{LLM}_{opt\text{-}4}$ & 66.69 & 45.89 & 49.09 & 69.39 & 48.20 & 52.01 \\
CoT-$\bm{LLM}_{opt\text{-}4}$ & \textbf{66.92} & \textbf{46.65} & \textbf{49.58} & \textbf{69.43} & \textbf{48.65} & \textbf{52.66} \\
\hline
\end{tabular}
\label{tab:ablation_study}
\end{table}
The use of both $\bm{LLM}_{opt\text{-}3.5}$ and $\bm{LLM}_{opt\text{-}4}$ significantly enhances the effectiveness of the initial prompting strategy, validating RecPrompt's ability to generate refined recommendation prompts. Notably, $\bm{LLM}_{opt\text{-}4}$ shows greater performance improvements over $\bm{LLM}_{opt\text{-}3.5}$, particularly when paired with $\bm{LLM}_{rec\text{-}3.5}$. This highlights the incremental benefits of employing more advanced LLMs as optimizers. Integrating category information results in significant improvements across all evaluated performance metrics, underscoring the importance of topical relevance in achieving accurate recommendations. This finding supports RecPrompt's strategic focus on leveraging $\bm{LLM}_{rec}$ to extract and summarize topics that reflect user interests and provide explanations for click behavior from a topical perspective.

\begin{figure}[!t]
    \centering
    \includegraphics[width=\linewidth, height=3.5cm]{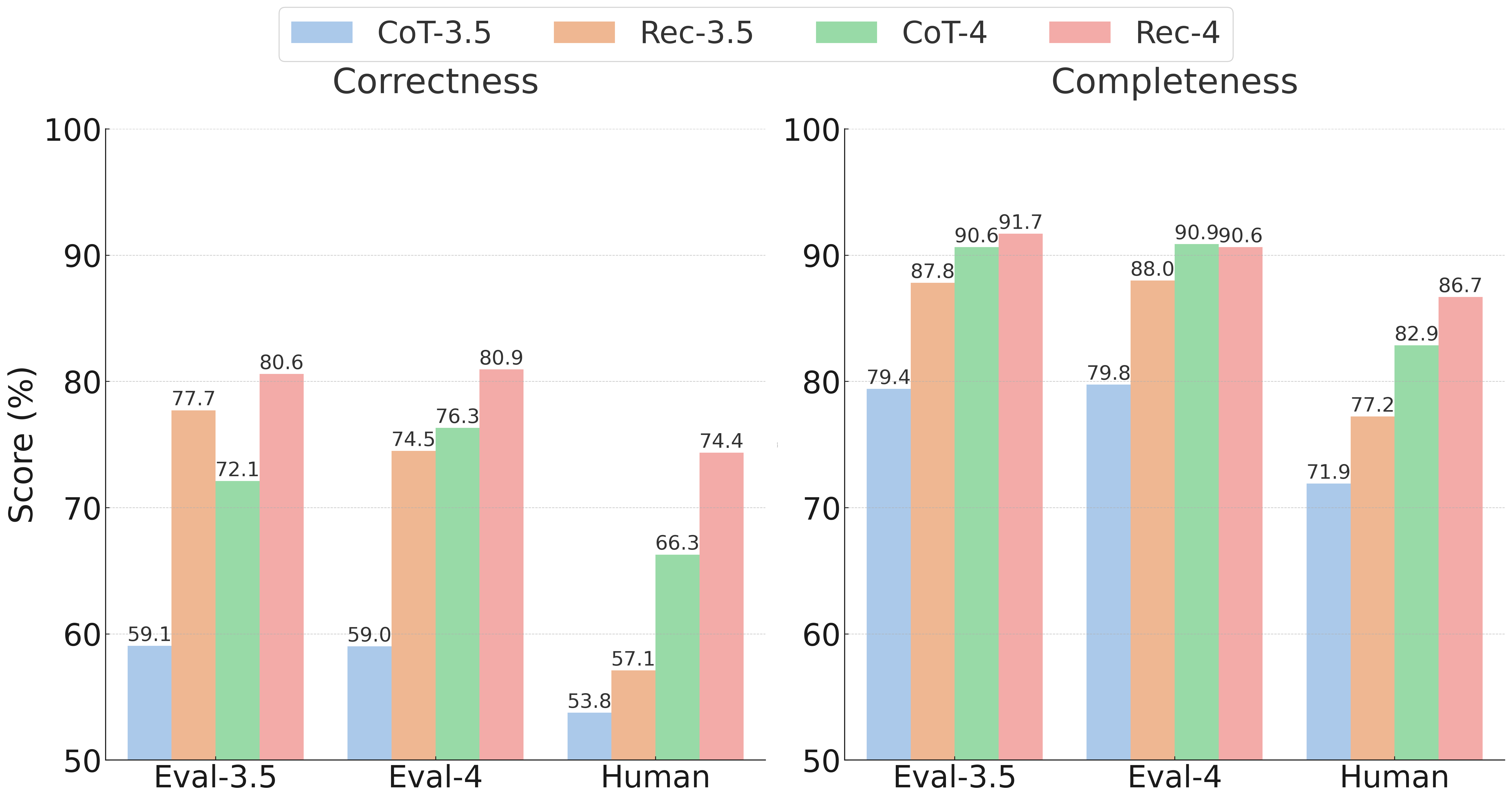}
    \caption{TopicScore evaluation between general CoT strategies and RecPrompt-CoT (Rec-3.5 and Rec-4), utilizing $\bm{LLM}_{rec\text{-}3.5}$ and $\bm{LLM}_{rec\text{-}4}$, optimized by $\bm{LLM}_{opt\text{-}4}$. }
    \label{fig:topic_score}
\end{figure}
\subsection{TopicScore Analysis}
In evaluating TopicScore, we used a dataset of 272 news articles and their associated 688 topics generated by $\bm{LLM}_{rec}$ under different settings. Due to substantial redundancy among the extracted topics, we consolidated similar topics, focusing on 214 unique topics for further evaluation. The evaluation metrics measured the precision of these topics in reflecting recommendations and their ability to encapsulate user interests comprehensively. 
We enriched our analysis by averaging TopicScore assessments from three human annotators and comparing them with LLM evaluations.
From Figure~\ref{fig:topic_score}, we observed only marginal differences in TopicScore evaluations between $\bm{LLM}_{eval\text{-}3.5}$ and $\bm{LLM}_{eval\text{-}4}$, with both generally assigning higher scores than human annotators. This suggests a potential discrepancy in the criteria for topic relevance assessment between machine and human evaluators. Notably, $\bm{LLM}_{rec\text{-}4}$ demonstrated a consistent advantage over $\bm{LLM}_{rec\text{-}3.5}$ in terms of topic accuracy and coverage, highlighting its enhanced capability for content analysis and summarization. 
Furthermore, the application of both $\bm{LLM}_{rec\text{-}3.5}$ and $\bm{LLM}_{rec\text{-}4}$ within the RecPrompt framework significantly elevated the TopicScore beyond the baseline established by the CoT strategy, validating RecPrompt's effectiveness in optimizing topic extraction and relevance.

\section{Conclusion}
This paper introduces RecPrompt, a novel framework using LLMs for news recommendation. It combines a recommender and a prompt optimizer with an iterative bootstrapping process to enhance prompting strategies. Our experiments show that RecPrompt significantly outperforms traditional and deep neural news recommendation models across various performance metrics.
Furthermore, we propose a novel metric, TopicScore, to assess the correctness and completeness of the topics generated by RecPrompt and compare them with the CoT prompting strategy. The results demonstrate the framework's ability to align recommendations closely with user interests and emphasize the critical role of prompt engineering in improving accuracy and user interest alignment when employing LLM-based recommendations. 

% \begin{acks}
% \end{acks}
\balance
\bibliographystyle{ACM-Reference-Format}
\bibliography{reference}
\clearpage

\end{document}